\documentclass{ws-ijmpa}

\begin{document}

\markboth{N. Ishibashi, Y. Baba and K. Murakami}
{D-branes and Closed String Field Theory}

%
\catchline{}{}{}{}{}
%

\title{D-BRANES AND CLOSED STRING FIELD THEORY}

\author{NOBUYUKI ISHIBASHI}

\address{Institute of Physics, University of Tsukuba,\\
Tsukuba, Ibaraki 305-8571, Japan
\\
ishibash@het.ph.tsukuba.ac.jp}

\author{YUTAKA BABA and KOICHI MURAKAMI}

\address{Theoretical Physics Laboratory, RIKEN,\\
         Wako, Saitama 351-0198, Japan\\
yutaka@het.ph.tsukuba.ac.jp and murakami@riken.jp }

\maketitle

\begin{history}
\received{Day Month Year}
\revised{Day Month Year}
\end{history}

\begin{abstract}
We construct solitonic states in the $OSp$ invariant string field 
theory, which are BRST invariant in the leading order of 
regularization parameter. 
One can show that these solitonic states describe D-branes 
and ghost D-branes, by calculating the scattering amplitudes. 

\end{abstract}

\ccode{PACS numbers: 11.25.Sq, 11.25.Uv}

\section{Introduction}	
D-branes have been studied for many years and used to reveal 
nonperturbative aspects of string theory. 
As A.~Sen argued, D-branes can be realized as soliton solutions 
in open string field theory. 
Now some of his conjectures\cite{Sen:1999mh}
have been proved analytically\cite{Schnabl:2005gv} 
in Witten's open string field theory.\cite{Witten:1985cc} 

What we would like to discuss here is how D-branes can be realized 
in closed string field theory. 
Hashimoto and Hata\cite{Hashimoto:1997vz} studied this problem 
in the context of HIKKO formulation.\cite{Hata:1986kj}
They modified the action by adding a source term made from 
the boundary state of a D-brane. 
They showed that such a source term does not break the gauge 
invariance, and argued that this term corresponds to the D-brane 
in the closed string field theory.  
Unfortunately one cannot fix the normalization of the boundary state 
only from the gauge invariance. 
Namely one cannot fix the tension of the D-brane in their formulation.

However for noncritical strings, the situation is better. 
D-branes in noncritical string theories 
can be defined as in the critical ones.\cite{Fateev:2000ik}
In Ref.~\refcite{Fukuma:1996hj}, Fukuma and Yahikozawa 
showed that the D-branes can be realized as solitonic operators 
which commute with the Virasoro 
and $W$ constraints\cite{Fukuma:1990jw} for the noncritical 
string theories.
In Ref.~\refcite{Hanada:2004im}, it was shown that how 
such solitonic operators are realized in the string field theory 
of noncritical strings presented in 
Ref.~\refcite{Ishibashi:1993pc}. 
States in which D-branes are excited can be made by acting 
these solitonic operators on the vacuum. 

What we would like to discuss in this article is a generalization of 
such a construction to the case of critical bosonic string theory. 
We will show that similar construction is 
possible\cite{Baba:2006rs}\cdash\cite{Baba:2007je} for 
the $OSp$ invariant string field 
theory.\cite{Siegel:1984ap}\cdash\cite{Kawano:1992dp}

\section{$OSp$ Invariant String Field Theory}
The $OSp$ invariant string field theory (SFT)
is basically a covariantized version of the 
light-cone gauge SFT.\cite{Kaku:1974zz} 
Therefore let us first briefly review the light-cone gauge SFT. 

\subsection{Light-cone gauge SFT}
\begin{figure}[htbp]
\begin{center}
\includegraphics[width=4cm,keepaspectratio=true]{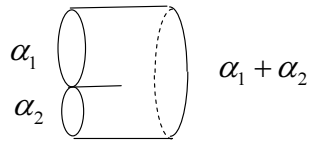}
\end{center}
\caption{Three-string interaction in the light-cone gauge}
\label{fig1}
\end{figure}

In the light-cone gauge SFT for closed strings, 
the string field $\Phi$ can be considered as a functional of 
the variables $t,\alpha ,X^i(\sigma )~(i=1,\cdots ,24)$, 
where $t=x^+$ and  $\alpha =2p^+$,
and $X^i$ are the coordinates in the transverse directions.  
As usual we consider a state $|\Phi (t,\alpha )\rangle$ 
in the Fock space of $X^i$ as a representative of the string field. 
The string field $|\Phi (t,\alpha )\rangle$ is taken to satisfy
$
\mathcal{P}|\Phi (t,\alpha )\rangle 
=
|\Phi (t,\alpha )\rangle 
$
, where
\begin{equation}
\mathcal{P}
=
\int_0^{2\pi}d\theta e^{i\theta (L_0-\tilde{L}_0)}. 
\label{projection}
\end{equation}
The action can be expressed as
\begin{equation}
S=
\int dt \left[
  \frac{1}{2}
   \int \frac{d\alpha}{4\pi}\alpha\left\langle \Phi (-\alpha ) \right|
       \left( i\frac{\partial}{\partial t}
                  -H
       \right)  \left|\Phi (\alpha )\right\rangle
+ \frac{2g}{3}
\int \frac{d\alpha}{4\pi}\alpha
\langle \Phi (-\alpha )|
\Phi \ast \Phi (\alpha )\rangle 
\right].
\label{action}
\end{equation}
Here $\langle \Phi (\alpha )|$ denotes the BPZ conjugate of 
$|\Phi (\alpha )\rangle$. 
$\ast$ denotes a product of the closed string fields $\Phi$, 
and describes the interaction of the strings depicted in 
Fig.~\ref{fig1}. 
The integration measure 
\begin{equation}
\int \frac{d\alpha}{4\pi}\alpha ,
\end{equation}
for $\alpha$ may look a little odd, but it yields the right kinetic 
term taking $\alpha =2p^+$ into account. 
The details of the notations are given in the appendix. 
The light cone gauge SFT possesses nonlinearly realized 
$SO(25,1)$ symmetry.  

\subsection{$OSp$ invariant SFT}
The $OSp$ theory can be obtained by considering the light-cone gauge 
SFT on a flat supermanifold  with extra coordinates 
$X^{25}, X^{26}, C, \bar{C}$ in addition to $X^i$. 
Here $C$ and $\bar{C}$ are Grassmann variables 
with conformal weight 0. 
The action for the $OSp$ theory is the same as eq.(\ref{action}),
but now the string field $|\Phi\rangle$ is in the Fock space of 
$C, \bar{C}, X^\mu$ $(\mu =1,\cdots ,26)$. 
One can prove that this theory possesses nonlinearly realized 
$OSp(27,1|2)$ symmetry, 
essentially because adding the extra variables does not change 
the Virasoro central charge. 

$X^\mu$ being regarded as Euclidean coordinates of 
the $26$ dimensional spacetime, 
the theory becomes a covariant formulation of the string theory. 
One can identify $C$ and $\bar{C}$ with the usual $bc$ ghosts 
in the conformal gauge. 
Therefore the $OSp$ invariant SFT can be considered as 
covariant SFT including extra variables $t,\alpha$, 
which correspond to time and length of the string. 

Since it is a covariant theory with ghost-like variables, 
we need a BRST symmetry. 
We regard the $C-$ component of the $OSp$ transformations 
as the BRST transformation. 
This is given as
\begin{equation}
\delta_\mathrm{B}|\Phi (\alpha )\rangle
=
\frac{1}{2}M^{C-}|\Phi (\alpha )\rangle
 -g \mathcal{P} C(\sigma_I)|\Phi \ast \Phi (\alpha )\rangle ,
\label{BRST}
\end{equation}
where $M^{C-}$ is the first quantized $OSp$ generator,
which is considered as the BRST charge, and 
$\sigma_I$ is the coordinate corresponding to the interaction point 
in the $\ast$-product. 
This BRST transformation is nilpotent by construction. 

It is easy to show that the string field Hamiltonian 
of the $OSp$ theory is BRST exact. 
Having an extra time and BRST exact hamiltonian, 
this theory cannot correspond to the usual formulation of field theory. 
It rather looks like that for stochastic quantization. 
Therefore we treat the theory in the way we do in such formulation. 
If one calculates Green's functions of BRST invariant observables, 
the results essentially depend only on the 26 dimensional coordinates 
and can be considered as Green's functions in a 26 dimensional theory. 
Then we can derive the S-matrix elements from these Green's functions 
in the usual way.  
The results can be proved to coincide with the S-matrix elements 
derived from the light-cone gauge SFT.\cite{Baba:2007je}
Thus we can use this $OSp$ theory to describe bosonic strings.

\section{D-brane States}
The formulation of the $OSp$ invariant SFT is different from 
that of the usual covariant SFT, 
but this theory has features in common with the noncritical SFT:
It involves string length variables, joining-splitting interaction 
and the extra time variable. 
Because of this similarity, 
it is conceivable that the construction of solitonic operator 
is also possible in this theory.  
As we will show in the following, 
we can construct second quantized BRST invariant states corresponding to D-branes, 
imitating the construction of the solitonic operator in the noncritical case. 

\subsection{Canonical quantization}
Let us quantize the $OSp$ invariant SFT, 
following the usual light-cone quantization. 
Decomposing the string field $|\Phi (\alpha )\rangle$ as
\begin{equation}
|\Phi (\alpha )\rangle 
=
|\psi (\alpha )\rangle +|\bar{\psi}(\alpha )\rangle ,
\end{equation}
where
\begin{eqnarray}
|\psi (\alpha )\rangle =|\Phi (\alpha )\rangle \theta (\alpha ), 
\qquad
|\bar{\psi}(\alpha )\rangle =|\Phi (\alpha )\rangle \theta (-\alpha ),
\end{eqnarray}
we consider $\psi$ as the annihilation operator 
and $\bar{\psi}$ as the creation operator. 
The canonical commutation relation can be given as
\begin{equation}
\left[
\, |\psi (\alpha )\rangle\, , \,
\langle \bar{\psi}(\alpha^\prime )| \, \right]
=
\frac{4\pi}{\alpha}\delta (\alpha +\alpha^\prime )\theta (\alpha )
\mathcal{P}.
\label{eq:ccr}
\end{equation} 
Then we define the second-quantized vacuum $|0\rangle \! \rangle $
which satisfies
\begin{equation}
|\psi (\alpha )\rangle |0\rangle \! \rangle =0.
\end{equation}
Thus states in the SFT can be made by acting the creation operators 
on this vacuum. 

\subsection{Boundary states}
In order to construct states describing D-branes, 
we need the boundary states corresponding to them. 
In the $OSp$ theory, 
we define the boundary state $|B\rangle$ as a state in the 
Fock space of $X^\mu ,C, \bar{C}$. 
They are taken to 
satisfy usual conditions for $X^\mu$, and Dirichlet conditions for 
$C$ and $\bar{C}$. 
(For the convention for the normalization of $|B\rangle$,
see Ref.~\refcite{Baba:2007tc} where $|B\rangle$ is
denoted by $|B_{0}\rangle$.)
Since we encounter divergences in the calculation, 
we modify $|B\rangle$ as
\begin{equation}
|B\rangle \rightarrow 
|B^\epsilon\rangle
\equiv
e^{-\epsilon H}
|B\rangle ,
\end{equation}
for $\epsilon\sim 0$, 
to use $|B^\epsilon\rangle$ as a regularization of $|B\rangle$. 
Since $H$ is BRST exact, it is a BRST invariant regularization. 

Now using these operators and state, 
we consider a state in the following form: 
\begin{equation}
|D \rangle \! \rangle
\equiv
\int d\zeta\, 
 \exp \left[ a
         \int^{0}_{-\infty} \frac{d\alpha}{4\pi}
            \, e^{\zeta \alpha}
         \langle B^\epsilon | \bar{\psi}(\alpha ) \rangle
         +F(\zeta ) \right]
|0\rangle\!\rangle ,
\label{D}
\end{equation}
where $a$ is a constant. 
Since $\bar{\psi}$ is the creation operator, 
this state has the effect of inserting boundaries in worldsheets. 
$a, F(\zeta )$ will be fixed by requiring that this state is 
BRST invariant.
The BRST variation of the state $|D \rangle \! \rangle$ can be 
calculated by substituting eq.(\ref{BRST}) 
and using the canonical commutation relations (\ref{eq:ccr}), 
and we obtain
\begin{eqnarray}
\delta_\mathrm{B}|D \rangle \! \rangle
&=&
\int d\zeta \,
\exp \left[ a
         \int^{0}_{-\infty} \frac{d\alpha}{4\pi}
            \, e^{\zeta \alpha}
         \langle B^\epsilon | \bar{\psi}(\alpha ) \rangle
         +F(\zeta ) \right]
\nonumber
\\
& &
\hspace{6mm}
\times
a\left\{
\int^{0}_{-\infty} \frac{d\alpha}{4\pi}
            \, e^{\zeta \alpha}
         \langle B^\epsilon |\frac{1}{2}M^{C-}
          | \bar{\psi}(\alpha )\rangle
\right.
\nonumber
\\
& &
\hspace{14mm}
-g
\int^{0}_{-\infty} \frac{d\alpha}{4\pi}
            \, e^{\zeta \alpha}
         \langle B^\epsilon |C(\sigma_I)
             |\bar{\psi}\ast\bar{\psi}(\alpha )\rangle
\nonumber
\\
& &
\hspace{14mm}
\left.
+ga
\int_{0}^{\infty} \frac{d\alpha}{4\pi}\alpha 
\langle \bar{\psi}(-\alpha )|
C(\sigma_I)
|\left(B^\epsilon (\zeta )\ast B^\epsilon (\zeta )\right)
(\alpha )\rangle
\right\}|0\rangle\!\rangle ,
\label{deltaB}
\end{eqnarray}
where 
\begin{equation}
|B^\epsilon (\zeta )\rangle
\equiv
|B^\epsilon \rangle 
    \frac{e^{-\zeta \alpha}}{\alpha}\theta (\alpha ).
\end{equation}

\subsection{Idempotency equations}
\begin{figure}[htbp]
\begin{center}
\includegraphics[width=4cm,keepaspectratio=true]{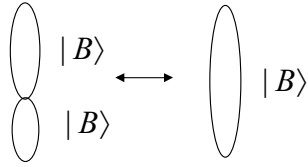}
\end{center}
\caption{Idempotency equation}
\label{fig2}
\end{figure}

In order to calculate the right hand side of eq.(\ref{deltaB}), 
we need the idempotency equations. 
The second term in the parenthesis $\{~\}$ corresponds to 
the process in which one boundary state splits into two, 
and the third term corresponds to the one 
in which two boundary states connect together. 
The states obtained from boundary states
connecting together or splitting into two 
should be expressed by the boundary states again,
as is described in Fig.~\ref{fig2}. 
Thus boundary states should satisfy so-called idempotency equation, 
which can be roughly written as
$B\ast B\propto B$.\cite{Kishimoto:2003ru}

In the case of the $OSp$ invariant SFT, 
one can obtain 
\begin{eqnarray}
& &
\int^{0}_{-\infty} \frac{d\alpha}{4\pi}
            \, e^{\zeta \alpha}
         \langle B^\epsilon |C(\sigma_I)
         |\bar{\psi}\ast\bar{\psi}(\alpha )\rangle
\nonumber
\\
&&
\hspace{1cm}
\sim
-2C_2
\int \frac{d\alpha}{4\pi} e^{\zeta \alpha}
\langle B|i\pi_0|\bar{\psi}(\alpha )\rangle 
\int \frac{d\alpha^\prime}{4\pi} \alpha^\prime e^{\zeta \alpha^\prime}
\langle B|\bar{\psi}(\alpha^\prime )\rangle ,
\nonumber
\\
& &
\int_{0}^{\infty} \frac{d\alpha}{4\pi}\alpha 
\langle \bar{\psi}(-\alpha )|
C(\sigma_I)
|\left(B^\epsilon (\zeta )\ast B^\epsilon (\zeta )\right)
(\alpha )\rangle
\nonumber
\\
& &
\hspace{1cm}
\sim
C_1\int \frac{d\alpha}{4\pi}\alpha
e^{\zeta \alpha}\langle B|i\pi_{0}|\bar{\psi}(\alpha )\rangle ,
\label{idempotency}
\end{eqnarray}
in the leading order in the regularization parameter $\epsilon$. 
Here $\pi_0$ is the momentum zero-mode for $C$ and 
$C_1$ and $C_2$ are constants given as
\begin{equation}
C_1
\equiv
\frac{(4\pi^{3})^{\frac{p+1}{2}} }{(2\pi)^{25}}
      \frac{4}{\epsilon^{2}(-\ln \epsilon)^{\frac{p+1}{2}}},
\qquad
C_2
\equiv
\frac{1}{(16 \pi)^{\frac{p+1}{2}}}
 \frac{4}{\epsilon^{2} (- \ln \epsilon)^{\frac{p+1}{2}}}.
\end{equation}

\subsection{States with one soliton}
Substituting eq.(\ref{idempotency}) into the right hand side of 
eq.(\ref{deltaB}), one can easily see that 
\begin{eqnarray}
&&
\delta_{\mathrm{B}}|D\rangle\!\rangle
\sim
ga^2 C_1
\int d\zeta 
\partial_\zeta 
\left(
\int \frac{d\alpha}{4\pi}
            \, e^{\zeta \alpha}
         \langle B |i\pi_0| \bar{\psi}(\alpha ) \rangle
\right.
\nonumber\\
&& \hspace{11em} 
\left. \times
\exp
\left[
a\int \frac{d\alpha'}{4\pi}
            \, e^{\zeta \alpha'}
         \langle B | \bar{\psi}(\alpha' ) \rangle
         +F(\zeta )
\right]
\right)
\, 
  |0 \rangle \! \rangle ,
\end{eqnarray}
provided
\begin{eqnarray}
F(\zeta )&=&b\zeta^2,
\nonumber \\
(a,b)&=&\pm (A,B),
\nonumber\\
A= \frac{(2\pi )^{13}}{(8\pi^2)^{\frac{p+1}{2}}\sqrt{\pi}},
\quad
&&
B=\frac{(2\pi )^{13}\epsilon^2(-\ln \epsilon )^{\frac{p+1}{2}}}
{16\left(\frac{\pi}{2}\right)^{\frac{p+1}{2}}\sqrt{\pi}g}.
\end{eqnarray}
Thus $\delta_{\mathrm{B}}|D\rangle\!\rangle =0$, 
if one takes the integration contour for $\zeta$ appropriately. 
$a$ and $F(\zeta )$ is fixed by the condition that the integrand 
becomes a total derivative. 

Therefore we have found two BRST invariant states
\begin{equation}
|D_{\pm}\rangle\!\rangle
\equiv
\int d\zeta
\exp
\left[\pm A
\int \frac{d\alpha}{4\pi}
            \, e^{\zeta \alpha}
         \langle B^{\epsilon} | \bar{\psi}(\alpha ) \rangle
 \pm B\zeta^2\right]
|0\rangle\!\rangle .
\end{equation}
It is possible to calculate scattering amplitudes perturbatively, 
using these states. 
One can show that we obtain the amplitudes in the presence of 
one D-brane using $|D_+\rangle\!\rangle$, if one takes $g>0$.  
Therefore $|D_+\rangle\!\rangle$ can be considered as a state with one 
D-brane. 
This result indicates that the boundaries are inserted
in the worldsheet with the right weight, 
and we have the right value for the tension of the D-brane. 
$|D_-\rangle\!\rangle$ may be considered as a state 
with one ghost D-brane.\cite{Okuda:2006fb} 

\subsection{States with multiple solitons}
One can generalize the above procedure and construct states with 
multiple D-branes. 
In order to do so, it is convenient to notice the following fact 
about the variable $\zeta$. 
Looking at the form of eq.(\ref{D}), one can see that 
boundaries on the worldsheet are inserted with the weight
\begin{equation}
\exp \left(-\zeta \times \mathrm{length~of~the~boundary}\right).
\end{equation} 
Thus $\zeta$ may be identified with the constant 
open string tachyon background. 
If there exist $N$ D-branes, it is natural to imagine that the 
variable $\zeta$ should become a hermitian $N\times N$ matrix $T$ and 
we should consider a state in the following form:
\begin{equation}
\int dT
 \exp \left[ a
         \int^{0}_{-\infty} \frac{d\alpha}{4\pi} \,
         \mathrm{Tr} \, e^{T \alpha} \,
         \langle B^\epsilon | \bar{\psi}(\alpha ) \rangle
         +\mathrm{Tr} \, F(T) \right]
|0\rangle\!\rangle .
\label{DN}
\end{equation}
Starting from eq.(\ref{DN}), 
one can proceed in the same way as above, 
just replacing $\zeta$ by $T$ and show that 
\begin{equation}
|D_{N,\pm}\rangle\!\rangle
\equiv
\int dT
\exp
\left[\pm A
\int \frac{d\alpha}{4\pi}
            \, \mathrm{Tr} \, e^{T \alpha} \,
         \langle B^{\epsilon} | \bar{\psi}(\alpha ) \rangle
\pm B \, \mathrm{Tr} \, T^2\right]
|0\rangle\!\rangle 
\end{equation}
are BRST invariant. 
It is easy to check that these states can be identified 
with the states with $N$ D-branes and ghost D-branes, 
by calculating the scattering amplitudes. 

\section{Conclusion}
We have constructed D-brane states in the $OSp$ invariant SFT 
for closed bosonic strings, as BRST invariant states. 
Imposing the condition that the states are BRST invariant, 
we can fix the value of the tension of the D-branes. 

There are many things to be pursued further. 
One thing is to consider similar construction for superstrings. 
Another thing is to study the relation between 
the variables $\zeta ,T$ and the open string tachyon further. 

The construction explained here may be useful to 
understand the dynamics of D-branes. 
For example, 
it may give some clue to Yoneya's ``trinity",\cite{Yoneya:2007ed} 
which is discussed by Yoneya in this conference.

\section*{Acknowledgments}
We would like to thank the organizers of the conference 
for arranging such a wonderful conference.  
This work was supported in part by Grant-in-Aid for
Young Scientists~(B) (19740164) from
the Ministry of Education, Culture, Sports, Science and
Technology (MEXT), and Grant-in-Aid for
JSPS Fellows (19$\cdot$1665).

\appendix

\section{Notations}
The notations employed here are different from those used in our 
original papers.\cite{Baba:2006rs}\cdash\cite{Baba:2007je} 
The string field $|\Phi\rangle$ here is a state in the Hilbert space 
of $X^{\mu}, C, \bar{C}$ including zero-modes. 
Therefore the inner product $\langle \Phi_1|\Phi_2\rangle$ here 
corresponds to 
\begin{equation}
\int d^{\prime\prime}r {}_r\langle \Phi_1|\Phi_2\rangle_r 
\end{equation}
in those papers, where
\begin{equation}
d^{\prime\prime}r
\equiv
\frac{d^{26}p_r}{(2\pi )^{25}}id\bar{\pi}^{(r)}_0d\pi^{(r)}_0.
\end{equation}
The $\ast$-product 
$\left(
\Phi_1\ast \Phi_2 
\right)
(\alpha)$ corresponds to the state
\begin{equation}
\frac{1}{\alpha} \int \frac{d\alpha_{1}}{4\pi}
\int d^{\prime\prime}1d^{\prime\prime}2
\, {}_{123}\langle 0|
e^{E(1,2,3)}\delta^{\prime\prime}(1,2,3)
|\Phi_1(\alpha_{1})\rangle_1|
\Phi_2(\alpha -\alpha_{1} )\rangle_2
\, |\mu (1,2,3)|^2,
\end{equation}
where $\alpha = -\alpha_{3}$ and 
\begin{equation}
\delta^{\prime\prime}(1,2,3)
=
(2\pi )^{25}\delta^{26}(p_1+p_2+p_3)
i(\bar{\pi}_0^{(1)}+\bar{\pi}_0^{(2)}+\bar{\pi}_0^{(3)})
(\pi_0^{(1)}+\pi_0^{(2)}+\pi_0^{(3)}) ,
\end{equation}
and the definitions of $E(1,2,3)$ and $|\mu (1,2,3)|^2$ are given in 
Refs.~\refcite{Baba:2007tc} and \refcite{Baba:2007je}. 
Notice that $\ast$ here is different from the $\ast$ 
appearing in those references. 


\end{document}